\documentclass[mathleft]{an}
\usepackage{graphicx}
\usepackage{times}
\begin{document}

\Pagespan{1}{10}
\Yearpublication{2006}%
\Yearsubmission{2005}%
\Month{1}%
\Volume{}%
\Issue{1}%

\title{X-Ray Observations of the Galactic Center with Suzaku}

\author{K. Koyama\thanks{Corresponding author: \email{koyama@cr.scphys.kyoto-u.ac.jp}\newline},  
Y. Hyodo,  T. Inui,  M. Nobukawa and  H. Mori}
\titlerunning{Instructions for authors}
\authorrunning{K. Koyama}
\institute{Department of Physics, Graduate School of Science, Kyoto University}

\received{30 May 2005}
\accepted{11 Nov 2005}
\publonline{later}


\abstract{
We report on the diffuse X-ray emissions from the Galactic center (GCDX) observed with the X-ray Imaging Spectrometer 
(XIS) on board the Suzaku satellite. The highly accurate energy calibrations and extremely low background of the XIS
provide many new facts on the GCDX. These are (1) the origin of the 6.7/7.0~keV lines is 
collisional excitation in hot plasma, (2) new SNR and super-bubble candidates are found, (3) most of the 6.4~keV line is
fluorescence by X-rays, and (4) time variability of the 6.4~keV line is found from the Sgr~B2 complex.
}

\maketitle

\section{Introduction}

The iron K-shell complex in the Galactic center diffuse X-rays (GCDX) is
composed of four lines at 6.4, 6.7, 7.0 and 7.1~keV (Koyama et al. 1986).  
These are K$\alpha$ lines of neutral(Fe\textsc{i}), He-like (Fe\textsc{xxv}) and H-like (Fe\textsc{xxvi}) iron, 
and K$\beta$ line of neutral iron (Fe\textsc{i}).
The origin of these iron lines or the origin of the GCDX has been a big problem.
In order to solve the problem, high quality data of the iron K-shell lines are essentially important.
Using the X-ray Imaging Spectrometer (XIS) (Koyama et al. 2007a) on board Suzaku (Mitsuda et al. 2007), 
we have obtained accurate X-ray spectra in the hard X-ray band including these key lines. 
This paper reports on the line analysis and discussion on these observational data. 

\section{Observations and Data Reduction}

The Galactic center regions were observed with the XIS (Koyama et al. 2007a) on board the Suzaku satellite
(Mitsuda et al. 2007). The observations are composed of several mosaic positions, each of $\sim$ 100~ks exposure.
We have made comprehensive energy calibrations on the whole imaging region of the CCDs on the XIS (for details, see Koyama et al. 2007d). 
The reliability of the energy scale is demonstrated in table 1. The K$\alpha$ lines of S\textsc{xvi},
Fe\textsc{xxvi} and Mn\textsc{i} (the calibration line) have simple structure compared 
to that of the K$\alpha$ complex of Fe\textsc{xxv}, and hence are reliably determined independent of emission mechanisms.  
As are given in table~1, the observed line energies of K$\alpha$ of S\textsc{xvi} and Fe\textsc{xxvi} are  within a few eV from 
the predicted 
energies of these lines.  Also Mn\textsc{i}K$\alpha$ is determined within a few eV 
from the laboratory (predicted) energy.  We therefore conclude that the energy scale is reliable within 5~eV in the 6--7~keV 
energy band. 

\begin{table}
\begin{center}
\begin{tabular}{llll} \\ 
\multicolumn{4}{l}{{\bf Table 1}: The observed and laboratory energies} \\
\hline \hline
H-like atom &  Observed (eV) &  Lab(eV) &  $\delta$(eV)\\
\hline
S\textsc{xvi}K$\alpha$  &  $2622^{+4}_{-3}$  &   2622   &     0\\
Fe\textsc{xxvi}K$\alpha$  & $6970^{+3}_{-2}$  &   6966    &   4\\
Mn\textsc{i}K$\alpha$  &  $5900^{+2}_{-1}$  &   5895   &    5 \\
\hline
\end{tabular}
\end{center}
\label{Table-cal}
\end{table}

\section{Origin of the 6.7~keV and 7.0~keV Lines }

The X-ray spectrum from the GC region exhibits many emi-
ssion lines.
The most prominent line in the low energy band is K$\alpha$ of S\textsc{xv} at
2.45 keV. In the high energy band, strong line-complex near at 6--7~keV is notable. These are
K$\alpha$ lines of  Fe\textsc{i}, Fe\textsc{xxv} and Fe\textsc{xxvi}, 
and K$\beta$ of Fe\textsc{i} at the energies of 
6.4, 6.7, 7.0 and 7.1~keV, respectively. 

\begin{figure}
\includegraphics[width=.45\textwidth,clip]{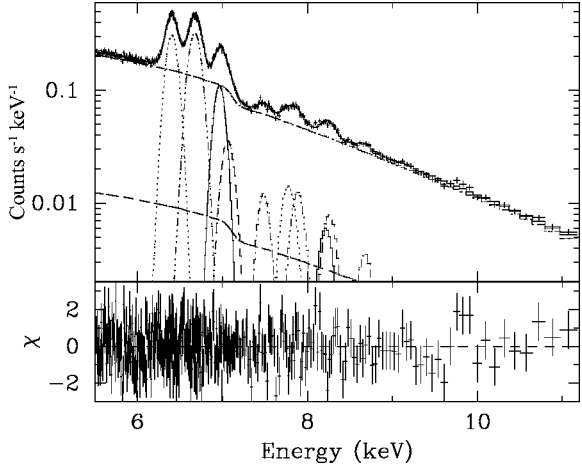}
\caption{The energy spectrum of the GCDX in the high energy band (adopted from Koyama et al. 2007d).
Four intense lines at 6--7 keV are K$\alpha$ of Fe\textsc{i}, Fe\textsc{xxv} and Fe\textsc{xxvi}, and K$\beta$ of Fe\textsc{i},
while weak lines at around 7--9 keV are K$\alpha$ of Ni\textsc{i} and Ni\textsc{xxvii}, and K$\beta$ of Fe\textsc{xxv}.
}
\label{GC-high}
\end{figure}

What is the origin of the 6.7~keV (K$\alpha$ of Fe\textsc{xxv}) line ? 
Is this due to collisional excitation (CE) or charge exchange (CX) ?
These two processes (CE and CX)  produce different line ratio of the resonance and forbidden lines 
in the fine structure of the K$\alpha$ line complex of Fe\textsc{xxv}, 
and hence the center energy of the 6.7~keV line is slightly different with each other. 
The laboratory experiments found that CX gives the center energy at 6666~eV, while CE is 
6680--6685~eV (Wargelin et al. 2005).
We made a very accurate spectrum of the GCDX in the hard energy  band up to $\geq$10~keV
(figure \ref{GC-high}). 
The observed line energy in the GCDX (figure 1) is determined to be $6680\pm1$~eV (systematic error is $\pm5$~eV), very close to CE of 6680--6685~eV
and significantly higher than that predicted by CX (6666~eV).

We detected narrow Fe\textsc{xxvi}K$\alpha$ (Ly$\alpha$) line at 7.0~keV.  
If the 7.0~keV line is due to CX, strong bump should appear at the Lyman series limit 
around 9--10~keV (the Lyman series transition of n$\geq 8\rightarrow$ n$=1$). 
The very low background of the XIS around 7--12 keV, the energy band above the iron K$\alpha$ complex, 
enable us to check the structure of the Lyman series limit.
We see no bump at 9--10~keV in figure \ref{GC-high}, which excludes CX origin.
Thus the origin of the 6.7 and 7.0~keV line is likely to be a CE process, or the X-rays are due to a high temperature plasma.

\begin{table}
\begin{center}
\begin{tabular}{lc} \\ 
\multicolumn{2}{l}{{\bf Table 2} The plasma temperatures determined from the line} \\
\multicolumn{2}{l}{flux ratios and the energy centroid of the 6.7~keV line} \\
\hline \hline
& Ionization  Temperature (keV) \\
\hline 
Fe\textsc{xxvi}K$\alpha$/Fe\textsc{xxv}K$\alpha$ & $6.5^{+0.1}_{- 0.1}$ \\
Fe\textsc{xxvi}K$\beta$/Fe\textsc{xxv}K$\beta$ & $5.1^{+1.5}_{-1.0}$ \\
Ni\textsc{xxviii}K$\alpha$/Ni\textsc{xxvii}K$\alpha$ & $9.3^{+1.6}_{-2.5}$ \\
Centroid of Fe\textsc{xxv}K$\alpha$ & $2.5-6.5$\\
\hline
&  Electron Temperature (keV)\\
\hline
 Fe\textsc{xxv}K$\beta$/Fe\textsc{xxv}K$\alpha$ & $6.2^{+3}_{-1}$ \\
 Fe\textsc{xxvi}K$\beta$/Fe\textsc{xxvi}K$\alpha$  & $\geq 6.5$ (lower limit)\\   
\hline 
\end{tabular}
\end{center}
\label{Table-temp}
\end{table}

We then discuss the plasma structure. Ionization temperature ($T_{\rm i}$) is determined by the ratio of K$\alpha$ 
lines from He-like and H-like atoms, while the electron temperature ($T_{\rm e}$) is determined 
by the ratio of K$\alpha$ and K$\beta$ lines from the same atom in the same ionization state. 
Since the Fe\textsc{xxvi}K$\alpha$ and Fe\textsc{xxv}K$\beta$ lines are near the iron K-edge, the fluxes are
strongly coupled to the depth of the K-edge. The depth of the K-edge structure can be precisely constrained
by the very low background data at 7--12~keV.
Figure 1 demonstrates that these line fluxes are determined very accurately due to the accurate and the 
low background flux level above the 7~keV band. 

The derived temperatures from the observed line flux ratios are given in table~2.
We note that electron temperature has been usually determined from the continuum shape assuming a 
thermal bremsstrahlung model. This method, however, may have large errors if the continuum is contaminated by a non-thermal
emission.
Our new method to determine the electron temperature by the ratio of K$\alpha$ and K$\beta$ lines from the same atom 
in the same ionization state,
is free from any contribution of a non-thermal component.

Since the energy centroid of the Fe\textsc{xxv}K$\alpha$ moves with the ionization temperature
due to the contamination of 
the satellite lines, we can also determine the temperature from the observed center energy. The result is also
given in table~2.
We see that all the results are consistent with the plasma of $\sim6.5$~keV temperature 
in collisional ionization equilibrium (CIE).

\section{Discovery of New SNRs}
\begin{figure}
\centerline{
\includegraphics[width=0.5\textwidth]{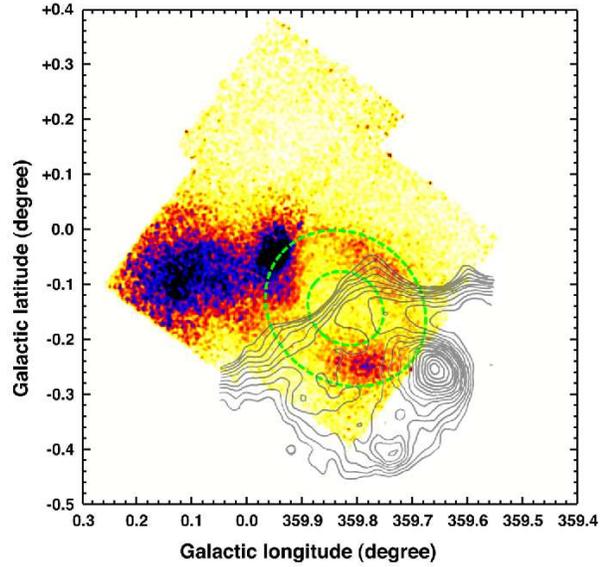}
}
\caption{The 2.45 keV (S\textsc{xv}K$\alpha$) band image overlaid on 
the  10~GHz band contour (light gray solid lines:Sofue 1988). 
The new X-ray SNR candidate G~359.79$-$0.26 is located at the center of the
radio shell. A large ring-like structure is shown  by the green-dashed annulus
(adopted from Mori et al. 2007).} 

\label{gr2.5red6.7bl6.4}
\end{figure}

\begin{figure}
\includegraphics[width=.45\textwidth, clip]{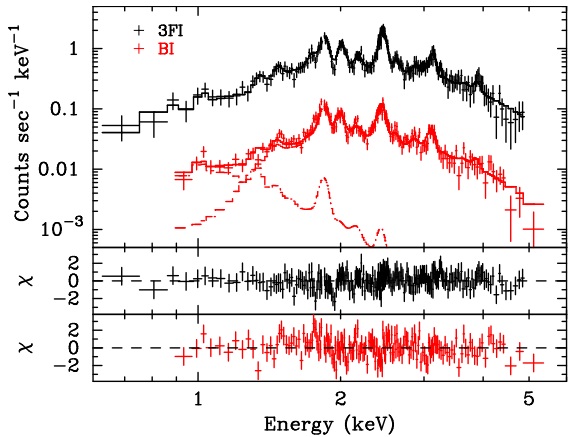}
\caption{The X-ray spectra of a  new SNR candidate, G~359.79$-$0.26 at the southeast of the Galactic center (adopted from Mori et al. 2007).}
\label{new-snr-1keV}
\end{figure}

Since the He-like K$\alpha$ lines are prominent in high temperature plasmas such as SNRs, we searched for SNR candidates 
using the Fe\textsc{xxv}K$\alpha$ (6.7~keV) and S\textsc{xv}K$\alpha$ (2.45~keV) lines. 
In the 6.7~keV band, we found two bright spots, Sgr~A~East (Koyama et al. 2007c) and G~0.61$+$0.01
(Koyama et al. 2007b).
The temperatures of these SNRs (3--4~keV) are hig-\\
her than any other young SNRs,
probably due to extreme environmental  conditions in the GC regions.

In the 2.45~keV band, we also found several bright spots, which may also be young SNRs of moderate 
temperature ($\sim$1~keV). All the SNR candidates have absorptions of (0.4--1)$\times10^{23}$cm$^{-2}$. These values are
the same as 
the interstellar absorption  toward the GC region, hence these SNR candidates are located near at the GC region.

We show a bright spot at $(l, b)=(359^\circ.79, -0.^\circ26$) (G~359.79$-$0.26: Mori et al. 2007) in figure 2.  
G~359.79$-$\\
0.26 is a strong candidate of SNR, because it is also associated with a radio ring structure (Sofue 1988).
The X-ray spectrum of G~359.79$-$0.26 is given in figure \ref{new-snr-1keV}. It is fitted with a thermal plasma model of $\sim$ 1-keV
temperature.

In addition to the SNR candidates,  we discovered large ring/arc structures.  In figure 2, we see a 
large ring with a 10 arcmin radius with the center at $(l, b)=(359^\circ.81, -0^\circ.14)$.  
G~359.79$-$0.26 is a part of this large ring. The real size of the ring is about
20-pc radius at the Galactic center distance of 8~kpc. Therefore this ring would be a supper-bubble. 
Another super-bubble candidate 
is a faint arc in the positive Galactic latitude emerging from Sgr~C ($l= 359^\circ.45$) to the positive Galactic latitude 
into the Galactic center with a radius of 30 arcmin. 
To confirm this faint structure, however, deep exposure observations are required.


\section{The 6.4~keV Clumps}

Figure 4 is the 6.4~keV line map. We see many clumps.
The spectra from these clumps exhibit not only a strong 6.4~keV line but also a 7.1~keV line, 
which are K$\alpha$ and K$\beta$ line from a neutral or low ionization Fe atoms.
The flux ratio is about 0.1, consistent with fluorescence from a neutral Fe (Fe\textsc{i}).

\begin{figure*}
\centerline{
\hspace*{10mm}
\includegraphics[width=0.8\textwidth]{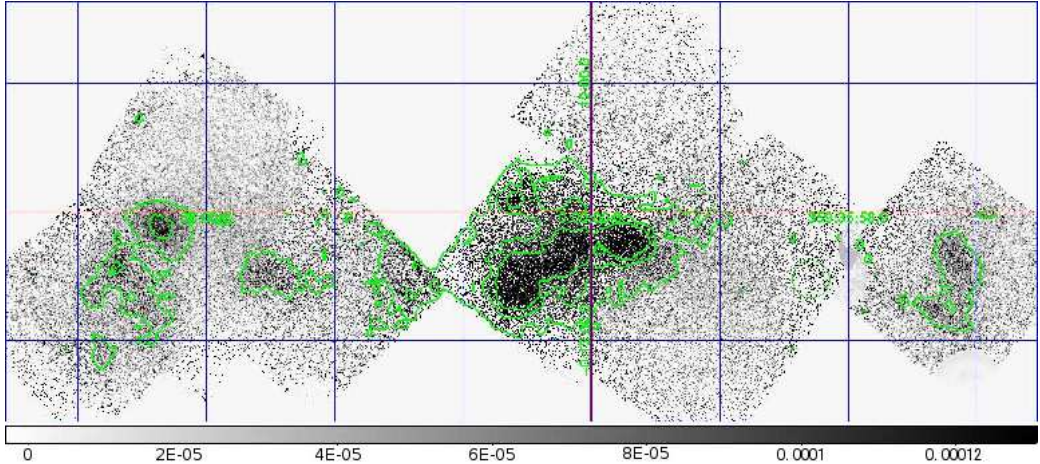}
}
\caption{The 6.4~keV (Fe\textsc{i}K$\alpha$) map. The grid is the Galactic coordinate with 
the spacing of 0.2 degree}
\label{6.4map}
\end{figure*}

What is the origin of the 6.4~keV clumps ? Is this due to inner shell ionization by electrons or X-rays ?
In table 3, we list the equivalent width ($EW$) of the 6.4 keV line and absorption depth ($N_{\rm H}$)
for the inner shell ionization by electrons and X-rays.
 From the spectrum analysis, we found that most of the 6.4~keV clumps have large $EW$ in 
the range  of 1--2~keV, which are
consistent with the fluorescence by X-ray irradiations (see table 3).
The K-edge absorptions ($N_{\rm H}$) are in the range of (2--10)$\times 10^{23} {\rm cm}^{-2}$, 
which are larger than the interstellar absorption to the GC. 
The large $N_{\rm H}$ is also consistent with the fluorescence by X-rays (see table 3).  
We, therefore conclude that most of the 6.4~keV clumps, if not all, are produced by X-ray irradiations. 

\begin{table}
\begin{center}
\begin{tabular}{lcc} \\ 
\multicolumn{3}{l}{{\bf Table 3} Inner shell ionization by electrons and X-rays}\\
\hline \hline
Parameter & Electrons & X-ray\\
\hline
$EW$ (keV) & 0.3--0.6 &  1--2 \\
$N_{\rm H}$ (cm$^{-2}$) & 10$^{21}$  & 	  10$^{24}$ \\
\hline
\end{tabular}
\end{center}
\label{Inner-shell}
\end{table}


\section{Time Variability of the Sgr~B2 complex}

A concrete fact to favor the X-ray irradiation origin would be time variability of the 6.4~keV line.
In order to examine the 6.4~keV line flux, we made the 
surface brightness map of the iron lines in the ASCA (1994) and Suzaku (2005) observations
(Koyama et al. 2007e, Inui et al. 2007) .
Since the energy resolution of ASCA was  not good enough to separate
the 6.4~keV line from the 6.7~keV line, we made images in the 6--7~keV band subtracting the continuum flux
(figure 5). 
We see clear decrease of the surface brightness from the ASCA to the Suzaku
observations.

\begin{figure}
\centerline{
\includegraphics[width=0.5\textwidth]{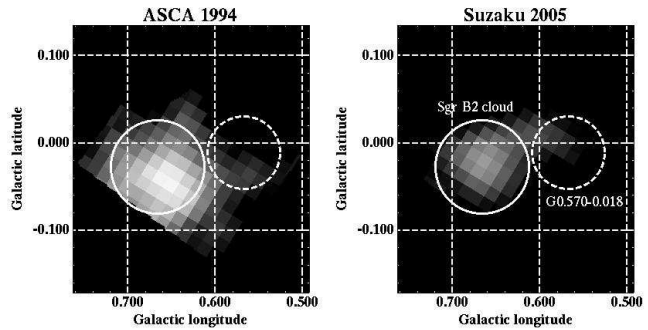}
}
\caption{The surface brightness maps of the 6.7 + 6.4~keV lines obtained with the ASCA SIS and the Suzaku XIS
(adopted from Inui et al. 2007).}
\end{figure}

\begin{figure}
  \begin{center}
  \includegraphics[width=0.45\textwidth]{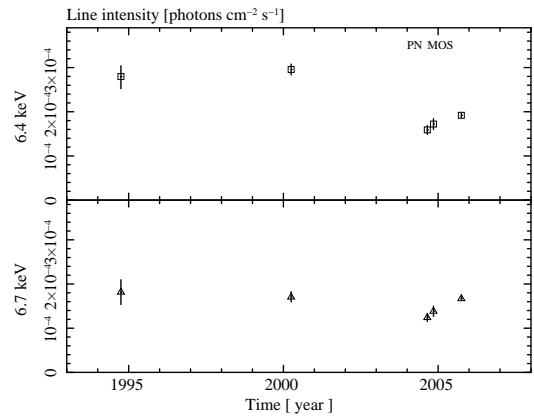}
  \end{center}
  \caption{The time history of the 6.4~keV and 6.7~keV line fluxes in the Sgr~B2 region
(adopted from Inui et al. 2007).}
  \label{fig:trend_b2area}
\end{figure}

In order to see the  flux change of the 6.4~keV line quantitatively, 
we made the X-ray spectra of the ASCA (1994), Chandra (2000), XMM-Newton (2004) and 
Suzaku (2005) observations near the Sgr~B2 cloud.
The spectra exhibit th-\\
ree pronounced peaks which represent 
Fe\textsc{i}K$\alpha$ (6.4~keV) and Fe\textsc{xxv}K$\alpha$ (6.7~keV) lines, 
and the composite of Fe\textsc{xxvi}K$\alpha$ (7.0~keV) and Fe\textsc{i}K$\beta$ (7.1~keV) lines.

Since the Suzaku spectrum  has the best statistics with accurate line energy (Koyama et al. 2007d),
we fit the Suza-\\
ku spectrum with a model of a power-law plus four Gaussian lines, fixing the energy
gap between Fe\textsc{i}K$\alpha$ (6400~eV)
and Fe\textsc{i}K$\beta$ (7058~eV) to the theoretical value (+658~eV) (Kaastra and Meve 1993).
The line flux ratio of Fe\textsc{xxvi}K$\alpha$/
Fe\textsc{xxv}K$\alpha$
is fixed to be 0.3. This ratio is slightly smaller than that found in the GCDX (Koyama et al. 2007d), 
because our selected region includes a SNR candidate G~0.61\\
$+$0.01, a strong 6.7~keV source (Koyama et al. 2007b).

Then we fit the spectra of the other satellites, fixing the power-law index, $N_{\rm H}$, the line center energies, and the 
flux ratio between the Fe\textsc{i}K$\alpha$ and Fe\textsc{i}K$\beta$ lines to those of the Suzaku best-fit parameters.
The best-fit fluxes of the 6.4~keV and 6.7~keV lines from all the satellites are given in figure 6 
as a function of observed years.

From figure 6, we see that the 6.7~keV line is almost constant. This is very reasonable
because the 6.7~keV line is due to the largely extended GCDX. In other words,
no time variability of the 6.7~keV line flux confirms the reliability of the cross-calibration
in the line flux of each satellite.
Based on this reliable cross-calibration, we firmly conclude that
the 6.4~keV line flux has been significantly variable, the fluxes in the XMM-Suzaku period
is 2/3 of that in the ASCA-Chandra era.

The linear size across the Sgr~B2 complex is about 40 light-years, but the brightest part (Sgr~B2) is $\sim$10 light-years.
The time scale of the 6.4~keV flux change is also $\sim$10 years, comparable to the light-crossing time
of the cloud. Any charged particle (with a finite mass) can not move as fast as the speed of light, and hence can not produce
such a rapid and large scale variability. 

A unique scenario to explain the spectral features and the fast time variability of the 6.4~keV lines
is that the Sgr\\
B2 cloud absorbs variable X-rays above the 7.1~keV edge energies and simultaneously 
re-emits the fluorescent 6.4\\
keV lines.  
Where is the bright variable X-ray source ? Is this a transient source located near 
the Sgr~B2 complex ?  This putative transient source must be brighter than 10$^{37}$~ergs~s$^{-1}$
for more than 10 years with the flux variability of a factor 1.5, which is very unlikely.

The most probable source to exhibit the bright and relatively long-lived X-rays is a massive black hole Sgr~A$^*$. 
About 300 years ago, the X-ray from Sgr~A$^*$  was 10$^6$ times brighter than the present value, 
and decayed to about 2/3 after $\sim$10 years.
The X-rays hit the Sgr~B2 cloud after $\sim$300
-years travel.
The cloud re-emits the 6.4~keV photons. Like a time delayed-echo, the X-ray is now just arriving at 
the Earth, when Sgr~A$^*$ becomes quiet.

\section*{Acknowledgements}
We would like to thank H. Nakajima, H. Uchiyama, and Y. Takikawa for their 
efforts in the Galactic center data analysis.
This work is supported by the Grant-in-Aid for the 21st
Century COE ``Center for Diversity and Universality in Physics'' from
Ministry of Education, Culture, Sports, Science and Technology (MEXT)
of Japan.

\end{document}